**Risks and opportunities in arbitrage and market-making in blockchain-based currency markets. Part 1 : Risks**

**Vittorio Astarita**
*vittorio.astarita@unical.it*

*https://www.researchgate.net/profile/Vittorio-Astarita*

*University of Calabria*
*Italy*

**Abstract**
This study provides a practical introduction to high-frequency trading in blockchain-based currency markets. These types of markets have some specific characteristics that differentiate them from the stock markets, such as a large number of trading exchanges (centralized and decentralized), relative simplicity in moving funds from one exchange to another, and the large number of new currencies that have very little liquidity. This study analyzes the possible risks that specifically characterize this type of trading operation, the potential opportunities, and the algorithms that are mostly used, providing information that can be useful for practitioners who intend to operate in these markets by providing (and risking) liquidity.

## 1. Introduction

Blockchain-based cryptocurrencies serve as the foundation for trading activities, where securities and commodities are exchanged 24 hours a day, every day of the year. This non-stop market activity takes place on more than 600 centralized exchanges and an 400 decentralized exchanges (according to Coingecko, 2023). As of April 2023, the total capitalization of blockchain-based currencies is approximately 1.23 trillion dollars (Coingecko, 2023). The dominant cryptocurrency, Bitcoin, which the American SEC considers a commodity, has a total market capitalization value of around 542 billion dollars (Wikipedia, 2023a).
Daily exchanged values range between 50 and 100 billion dollars. However, this figure may not be entirely reliable because many exchanges or issuers engage in wash trading to inflate the volumes exchanged artificially. Wash trading is a prohibited practice on stock markets, where a single entity creates trades by simultaneously selling and buying the same financial instruments. This creates a false impression of market activity without incurring market risk or changing the entity's market position (Wikipedia, 2023b).

One of the problems in this market is the lack of regulations. Centralized exchanges (CEX) are localized worldwide, sometimes to avoid specific regulations, and decentralized exchanges (DEX) supposedly do not even have a defined location. The number of blockchain-based financial instruments that are traded is stunning, according to Coingecko in April 2023 around 10800 "coins" are traded on CEX and DEX platforms (Coingecko, 2023).

## 2. Risk and opportunities for traders

Traders have the opportunity to trade on more than 1000 exchanges with more than 10000 financial instruments (Coingecko, 2023). Each exchange also lists different trading pairs for each financial instrument, creating an incredible number of markets in which to trade. Differences in prices arise naturally between different trading pairs traded on different platforms (especially when volatility is high or, in simpler words, when the price is changing rapidly with time), creating opportunities for trading practices such as arbitrage and market making. Traders can turn the provision of liquidity into profit by applying the techniques described in the next section. Moreover, the whole market seems to be constantly growing following the Bitcoin growing price, so the entire environment seems very favorable for traders. Unfortunately, this market is highly competitive and full of dishonest operators.

The average crypto-trader faces several potential pitfalls. The primary risks include:

1)      The risk of investing in and holding blockchain-based currencies without understanding the underlying economic details and technical aspects (Investopedia, 2023a).
2)      The risk of malpractice when handling financial instruments in self-managed wallets (Cointelegraph, 2023).
3)      The risk of transferring funds on blockchains using incorrect parameters, which can result in the loss of transferred funds (Coindesk, 2023).
4)      The risk of leaving funds on decentralized exchanges (DEX) and centralized exchanges (CEX), which can be hacked or fraudulently administered, leading to permanent loss of funds.
5)      The risk associated with trading operations.

While the first three risks can be mitigated through proper education in computer science and understanding how blockchain-based currencies work, the fourth and fifth risks cannot be entirely avoided. A discussion on reducing the fifth risk is presented in the next section, as it is connected with the adopted trading strategy.

## *2.1 Investing in and holding blockchain-based currencies without understanding the economic details and technical aspects of the chosen financial instrument.*

Investing in and holding blockchain-based currencies without a comprehensive understanding of the economic details and technical aspects exposes traders to significant risks. These risks include holding financial instruments that can suddenly decrease in price unexpectedly.

### *2.1.1 Blockchain-based currency risk*

Some blockchain-based currencies may appear to be decentralized, but they are actually controlled by single entities that hold the right of issuance or possess a majority of the funds since the initial emission. In the first scenario, controlling entities can dilute the supply of a financial instrument by introducing undesired inflation. These entities have numerous methods to execute such scams, including altering the rules of emission or exploiting existing regulations. For instance, some blockchain-based currencies do not have a maximum number of coins that will be generated, or they may have vague rules for coin emissions and inflation regulations that prove to be disadvantageous for smaller investors (Nakamoto, 2008).

### *2.1.2. Desiderable technical details for chosen financial instruments*

The following currency features could help protect investors from potential losses:

 **-Limited supply:** A currency with unlimited supply is susceptible to inflation. Inflation can debase a currency, reducing its value in a short time. In contrast, a limited and well-established maximum supply guarantees that a currency cannot be debased (Nakamoto, 2008).
        Decentralized governance and transaction verification process: Centralized governance and/or centralized blockchain ledger can expose the currency to changes in supply rules, arbitrary new rules introduction, or potential frauds affecting smaller investors (Mazieres, 2015).

 **-Clear monetary policy:** A transparent monetary policy that is easy to understand helps protect people from risks associated with obscure policies where variable rules can be imposed. Fixed rules established by mathematical algorithms can offer some protection but do not entirely solve this issue. For example, the case of the Terra Luna system, where the UST stablecoin was algorithmically linked to the LUNA coin for stabilization, demonstrates this problem. The UST stablecoin eventually lost its peg to the dollar value because the total value of UST was no longer backed by LUNA. As the UST stablecoin began to lose its peg, people who held the stablecoin started selling it off, losing confidence in the system. This led to arbitrageurs exploiting the algorithm that was supposed to keep the value of UST pegged to the dollar by buying UST on exchange open markets at a discount over USD value and transferring it to Terra Station (Kwon & Shin, 2018). On Terra Station, it was possible to redeem UST at parity with the dollar for LUNA coins. Once returned to exchanges, LUNA coins were sold at market value. This arbitrage was carried out for a couple of days, bringing the value of LUNA and UST close to zero. The Terra-Luna system was based on algorithmic rules that were potentially unstable under certain conditions, creating a self-feeding loop of debasement that was not foreseen, even by experienced investors (Kwon & Shin, 2018). Algorithmic rules and math can support investors, but they cannot guarantee a risk-free investment.

-**Open and permissionless:** When a currency is governed by self-appointed policymakers, there is a risk of losing access to funds. A reliable currency should always be usable and fungible, and anyone, anywhere, should have the ability to use it (Nakamoto, 2008). Currencies where controlling parties can exert influence and control to restrict access, limit usage (and even reverse and/or block transactions) should be avoided, as smaller investors might suffer the consequences sooner or later.

## 2.2 Risk of malpractice holding financial instruments in self-managed wallets.

Self-managed wallets were introduced with the aim of liberating users from intermediaries that would hold funds for investors in their custody. The rationale behind this innovation is that intermediaries such as exchange businesses, banks, and investing firms could misuse funds, lose them, or, as has occurred in these markets many times, be hacked (or pretend to be hacked) and lose access to client funds. Self-custody with dedicated software or paper wallets would remove the risks discussed in the next section. Unfortunately, this is not always the case, as self-custody requires an understanding of how these digital instruments work. This understanding is difficult for many users.

The main risks of self-managed custody are the permanent loss of the secret spending key and the loss of spending keys due to theft or computer hacks. Keys stored on a computer connected to the internet are only as secure as the computer itself. Physical conservation of keys on unconnected devices is also risky, as cold storage devices can be lost or destroyed. Memorizing keywords exposes users to the risk of forgetting them.

The best way to mitigate the risk of being hacked on the computer hosting the wallet is to hold secret keys in cold storage using an air-gapped device to sign transactions. However, this is not feasible for all blockchain-based instruments, as not all financial instruments have easy-to-use wallet software for offline signing of transactions.

## 2.3 Transfering funds on blockchains using the wrong parameters with the result of losing the transferred funds.

Transferring funds on blockchains using incorrect parameters is a common mistake that can happen due to a lack of experience or simple distraction (Medium, 2023). This error can lead to the use of the wrong destination address and the permanent loss of funds (Chainalysis, 2023). It's essential to understand the various aspects of transferring funds on blockchains, as well as the potential risks involved, to avoid such mistakes.

When transferring funds, it's crucial to use the correct blockchain transmission protocol. Different blockchains may have unique protocols and address formats, making it easy to accidentally send funds to an incompatible address (Chainalysis, 2023). . For instance, sending Bitcoin to an Ethereum address, or vice versa, will result in the loss of those funds. To avoid this, it's essential to double-check the destination address and ensure it matches the appropriate blockchain protocol.

Another potential issue that could lead to the loss of funds is the use of outdated or unsupported software (Investopedia, 2023b). Wallets or exchanges that haven't been updated to the latest protocol version may not recognize newer address formats, causing transactions to fail or funds to be lost (Investopedia, 2023b). Always ensure that the software being used is up-to-date and compatible with the intended blockchain network.

Human error can also play a significant role in the loss of funds during transfers (Chainalysis, 2023). Typing errors, copy-pasting mistakes or misreading an address can result in sending funds to an incorrect or non-existent address (Chainalysis, 2023). To minimize the risk of human error, always double-check the entered information and consider using QR codes, if available, to input the destination address.

In summary, transferring funds on blockchains using the wrong parameters can lead to the permanent loss of funds (Chainalysis, 2023). To prevent this, it's essential to understand the specificities of the blockchain protocol being used, ensure the use of up-to-date and compatible software, and take extra care when inputting destination addresses. By following these best practices, users can reduce the risk of losing funds during blockchain transfers.

**2.4. The risk of leaving funds on DEX and CEX, which can be hacked or fraudulently administrated, with consequent permanent loss of funds.**

Trading digital assets on decentralized exchanges (DEX) and centralized exchanges (CEX) involves inherent risks, including the potential for hacking, fraudulent administration, and the permanent loss of funds. The author of this text has personally experienced these risks, having been involved in the recent Polarity DEX hack, which resulted in the potential loss of around $500,000 of funds. This serves as a stark reminder of the consequences that can arise from such incidents. While DEX platforms claim to be exempt from such risks, many still have centralized points of entry and exit, making them susceptible to similar threats as CEX (Putz and Pernul, 2019).

Several recent cases of DEX hacks have led to significant fund losses, such as Hedera (March 9, 2023), Platypus (February 17, 2023), and Orion Protocol (February 3, 2023) [3]. According to ChainSec, as of March 2023, there have been a total of 132 DeFi exploits, with lost funds amounting to approximately $4.2 billion [4]. The largest of these exploits was the Ronin bridge hack, which resulted in the loss of $625 million in value.

CEX platforms have also experienced significant losses of user funds, with the infamous Mt. Gox hack in February 2014 being one of the earliest and most prominent examples (Decker and Wattenhofer, 2014). The risks associated with both DEX and CEX platforms highlight the importance of carefully considering fund storage and trading practices (Fantazzini and Calabrese, 2021).

To mitigate these risks, investors may consider using hardware wallets or cold storage solutions for holding long-term investments, while only maintaining a small amount of funds on exchanges for active trading. Additionally, it's essential to stay informed about the security features and reputation of the exchanges being used, as well as to enable two-factor authentication (2FA) to provide an additional layer of security.

To help individuals navigate the cryptocurrency market, resources like HedgewithCrypto offer helpful guides, unbiased reviews, and comparisons of various crypto exchanges and platforms (Hedgewithcrypto, 2023). Founded in June 2019, HedgewithCrypto has reached over 1.6 million readers worldwide, aiming to help them get started in the cryptocurrency market without having to worry about getting overwhelmed or scammed (Hedgewithcrypto, 2023).

To help individuals navigate the cryptocurrency market, resources like HedgewithCrypto offer helpful guides, unbiased reviews, and comparisons of various crypto exchanges and platforms (Hedgewithcrypto, 2023). Founded in June 2019, HedgewithCrypto has reached over 1.6 million readers worldwide, aiming to help them get started in the cryptocurrency market without having to worry about getting overwhelmed or scammed. According to HedgewithCrypto a cumulative amount of nearly USD $2.72 Billion has been stolen from crypto exchanges since 2012, with one of the latest being : FTX, which lost USD 600 million in a recent hack. Moreover, it states that approximately 0.3% of the total market cap (904 billion) has been lost to crypto exchange hacks and at least 47 Bitcoin exchanges have lost funds through a major cybersecurity breach. This is the list :

| Date | Exchange | Cause of Hack | Amount Stolen (USD) |
|---|---|---|---|
| 022, November 12 | FTX | Unauthorized transactions | $600 million |
| 2022, January 17 | Crypto.com | Unknown$97 million | $34 million |
| 2021, December 11 | AscendEX | Obtained access to hot wallet | $80 million |
| 2021, December 5 | BitMart | Obtained access to hot wallet | $150 million |
| 2021, August 19 | Liquid | Obtained access to hot wallet | $97 million |
| 2021, April 29 | Hotbit | Obtained access to hot wallet | Nil |
| 2020, December 23 | Livecoin | Compromised system/servers | Unknown |
| 2020, December 21 | EXMO | Obtained access to hot wallet | $4 million |
| 2020, December 1 | BTC Markets | Internal staff error/mistake | 270,000 user's private details |
| 2020, September 25 | KuCoin | Data leak | $275 million |
| 2020, July 11 | Cashaa | Malware | $3.1 million |
| 2020, June 29 | Balancer | Vulnerability in protocol | $500,000 |
| 2020, April 19 | Lendf.me | Bugs and Re-entrancy attack | $24.5 million |
| 2020, April 19 | Uniswap | Bugs and Re-entrancy attack | $500,000 |
| 2020, February 5 | Altsbit | Obtained access to hot wallet | $70,000 |
| 2019, December 19 | Youbit | Obtained access to hot wallet | Unknown |
| 2019, November 26 | Upbit | Obtained access to hot wallet | $49 million |
| 2019, November 5 | Vindax | Unknown | $500,000 |
| 2019, July 11 | Bitpoint | Compromised system/servers | $32 million |
| 2019, June 27 | Bitrue | Compromised system/servers | $4.5 million |
| 2019, June 6 | Gatehub | Unknown | $9.5 million |
| 2019, May 7 | Binance | Obtained access to hot wallet | $40 million |
| 2019, March 29 | Bithumb | Unknown | $29 million |
| 2019, March 25 | Coinbene | Suspected trusted insider | $40 million |
| 2019, March 24 | DragonEX | Unknown | $1 million |
| 2019, February 15 | Coinmama | Data leak | 450,000 user's private details |
| 2019, February 1 | Cryptopia | Unknown | $16 million |
| 2019, January 26 | LocalBitcoins | Phishing data on fake site | $27,000 |
| 2018, October 28 | Maplechange | Suspected trusted insider | $97 million |
| 2018, September 14 | Zaif | Obtained access to hot wallet | $60 million |
| 2018, June 18 | Bithumb | Unknown | $31 million |
| 2018, June 10 | Coinrail | Unknown | $40 million |
| 2018, April 13 | CoinSecure | Suspected trusted insider | $3.5 million |
| 2018, February 10 | Bitgrail | Suspected trusted insider | $146 million |
| 2018, January 27 | CoinCheck | Unknown | $560 million |
| 2017, December 20 | EtherDelta | Server DNS compromised | $1.4 million |
| 2017, July 5 | Bithumb | | $7 million |
| 2017, April 22 | Yapizon | Obtained access to hot wallet | $5.3 million |
| 2016, August 2 | Bitfinex | Unknown | $623 million |
| 2016, May 9 | Gatecoin | Obtained access to hot wallet | $2.14 million |
| 2016, April 7-9 | Shapeshift | Suspected trusted insider | $230,000 |
| 2016, February 16 | BTER | Unknown | $1.75 million |
| 2015, December 11 | Bitstamp | Malware | $5 million |
| 2015, August 15 | BTER | Suspected trusted insider | $1.65 million |
| 2014, July 13 | Mintpal | Obtained access to hot wallet | $2 million |
| 2014, March 4 | Poloniex | Obtained access to hot wallet | $50,000 |
| 2014, February | Mt. Gox | Various methods | $460 million |
| 2013, November 11 | Bitcash | Compromised system/servers | $100,000 |
| 2012, September 12 | Bitfloor | Compromised system/servers | $250,000 |
| 2012, March 1 | Bitcoinica | Compromised system/servers | $87,000 |

**Table 1.** List of comprised CEX according to : https://www.hedgewithcrypto.com

In summary, both DEX and CEX platforms come with inherent risks, including hacking and fraudulent administration, which can lead to the permanent loss of funds. To minimize these risks, investors should take precautions such as using cold storage solutions, maintaining minimal funds on exchanges, and staying informed about the security features of the platforms they use.

### 3. Operating trading risks.

Operating trading risks and trading algorithms that can minimize the risk are discussed in part 2 of this work.

### 4. Conclusion

This study provides a practical introduction to high-frequency trading in blockchain-based currency markets. This study presents the possible risks that specifically characterize this type of trading operation showing that opportunities may be completely erased by the risks of these new financial instruments. Part 2 will complete this draft work. This work does not encourage or promote trading activities. These kind of financial instruments are new and based on advanced technical concepts which require a proper understanding. Moreover even a proper understanding of risks is not enough to completely avoid them, especially the risk of failing intermediaries presented in section 2.4.